\documentclass[twocolumn,showpacs,preprintnumbers,superscriptaddress]{revtex4}
\usepackage{graphicx}
\usepackage{dcolumn}
\usepackage{bm}
\usepackage{color}
\usepackage{amsmath}
\usepackage{amssymb}
\usepackage{amsfonts}
\usepackage{esint}

\begin{document}

\title{Quantum entanglement distribution with hybrid parity gate}

\author{Feng Mei}
\affiliation{Laboratory of Photonic
Information Technology, SIPSE $\&$ LQIT, South
China Normal University, Guangzhou 510006, China}
\author{ Ya-Fei Yu}
\affiliation{Laboratory of Photonic
Information Technology, SIPSE $\&$ LQIT, South
China Normal University, Guangzhou 510006, China}
\author{ Xun-Li Feng}
\affiliation{Laboratory of Photonic
Information Technology, SIPSE $\&$ LQIT, South
China Normal University, Guangzhou 510006, China}
\affiliation{Centre for Quantum Technologies and Department of Physics,
National University of Singapore, 3 Science Drive 2, Singapore
117543, Singapore}
\author{Zhi-Ming Zhang}
\email{zmzhang@scnu.edu.cn}
\affiliation{Laboratory of Photonic
Information Technology, SIPSE $\&$ LQIT, South
China Normal University, Guangzhou 510006, China}
\affiliation{Centre for Quantum Technologies and Department of Physics,
National University of Singapore, 3 Science Drive 2, Singapore
117543, Singapore}

\author{C. H. Oh}
\affiliation{Centre for Quantum Technologies and Department of Physics,
National University of Singapore, 3 Science Drive 2, Singapore
117543, Singapore}

\date{\today}

\begin{abstract}
We propose a scheme for entanglement distribution among different single
atoms trapped in separated cavities. In our scheme, by reflecting an input
coherent optical pulse from a cavity with a single trapped atom, a
controlled phase-shift gate between the atom and the coherent optical pulse
is achieved. Based on this gate and homodyne detection, we construct an $n$%
-qubit parity gate and show its use for distribution of a large class of
entangled states in one shot, including the GHZ state $\left\vert
GHZ_{n}\right\rangle $, W state $\left\vert W_{n}\right\rangle $, Dicke
state $\left\vert D_{n,k}\right\rangle $ and certain sums of Dicke states $%
\left\vert G_{n,k}\right\rangle $. We also show such distribution could be
performed with high success probability and high fidelity even in the
presence of channel loss.
\end{abstract}

\maketitle

\section{Introduction}

In quantum information science, quantum networks play an important role,
which can accomplish tasks that are impossible in the realm of classical
physics. Quantum networks are composed of quantum nodes for storing and
processing information and quantum channels that link the nodes {\color{blue}%
\cite{QN}}. Distributing entanglement over different nodes through quantum
channels is a critical requirement, which could allow physical
implementation of quantum cryptography {\color{blue}\cite{QS1,QS2}}, quantum
secret sharing {\color{blue}\cite{QSS} }and distributed quantum computation
(QC) {\color{blue}\cite{DQC}}. It has also been found applications in
quantum metrology {\color{blue}\cite{PM}} and simulating quantum many-body
physics {\color{blue}\cite{QSM}}. Many protocols have been proposed for such
distribution, including the distribution between two distant single atoms {%
\color{blue}\cite{Atom1,Atom2}}, ions {\color{blue}\cite{Ion}, }atomic
ensembles {\color{blue}\cite{DLCZ} }and NV centers {\color{blue}\cite{NV}}.
These distributed protocols have high fidelities, but they all rely on
postselection by using single-photon detection with very low success
probabilities.

An alternative way is to use coherent light instead of single-photon as
quantum channels. This way also uses postselection but with homodyne
detection of a coherent light. Because the homodyne detection could be done
with near-unit efficiency {\color{blue}\cite{HD}}, so this distributing way
has a high success probability. It has been proposed recently to realize a
quantum repeater {\color{blue}\cite{CL,CL2}}. In the proposal, through a
bright coherent optical pulse sequentially interacting with two electronic
spins placed in cavities, entanglement can be distributed between two
distant spins by a homodyne detection of the final outgoing coherent pulse,
with the fidelity determined by the distinguishability $d\approx \alpha
\theta $ {\color{blue}\cite{CL,CL2}}. However, the nonlinear phase shift $%
\theta $ of the coherent light pulse generated by the dispersive
light-matter interaction in {\color{blue}\cite{CL,CL2} }is very small $(\sim
10^{-2})$. Considering the saturation effect, the fidelity can be improved
to modest level at most by increasing the amplitude of the coherent light $%
\alpha $. Moreover, this proposal only can project the two distant spins
into a Bell state with a probability of success $1/2$. In this paper, with
single atoms trapped in separated cavities as quantum nodes and a coherent
light as quantum channel, we propose a scheme for entanglement distribution
among two different nodes by a two-qubit parity gate. Because the parity
gate in our scheme relies on both the discrete atomic qubit state and the
continuous quantum variable of coherent light amplitude, we call it hybrid
parity (HP) gate. The basic building block of the HP gate is a local
controlled phase-shift (CPS) gate between a single atom trapped in a cavity
and an input coherent light. Based on two local CPS gates and a homodyne
detection of the final output coherent optical pulse, one can achieve a
two-qubit HP gate. Compared with the proposal {\color{blue}\cite{CL,CL2}},
the nonlinear phase shift $\theta $ of the coherent light in the CPS gate is
huge and can approach the order of $\pi $, which means that the entanglement
distribution in our scheme is with high fidelity. Besides, the two distant
atoms can be projected into a Bell state in one shot using the HP gate. It
is also found that the phase shift could be tuned to realized a $n$-qubit HP
gate.

With such $n$-qubit HP gate, we can distribute a large class of multiqubit
entanglement in one shot, like $n$-qubit GHZ state $\left\vert
GHZ_{n}\right\rangle $, W state $\left\vert W_{n}\right\rangle $, Dicke
state $\left\vert D_{n,k}\right\rangle $, and certain sums of Dicke states $%
\left\vert G_{n,k}\right\rangle $. These states are essential in some
quantum information processing (QIP) protocols, which cover quantum dense
coding {\color{blue}\cite{QDC}}, quantum key distribution {\color{blue}\cite%
{QKD}}, quantum teleportation {\color{blue}\cite{QT}}, quantum telecloning {%
\color{blue}\cite{QTC}}, quantum secret sharing ($\left\vert
G_{n,k}\right\rangle $) {\color{blue}\cite{QSS2} }and scalable quantum
search ($\left\vert D_{n,n/2}\right\rangle $) {\color{blue}\cite{Search}}.
We also show that our scheme inherits both the advantages of the single
photon and coherent light quantum channel based entanglement distribution
protocols and could be performed with a high fidelity and high success
probability. In addition, compared with the earlier QC protocols using
cavity input-output process {\color{blue}\cite{DK,LE,XX,Mei}}, our scheme
does not require high-$Q$ cavity working in the strong coupling regime and
not employ single-photon source, which greatly relaxes the experimental
requirement.

This paper is organized as follows. In Sec. II we concretely describe the
basic building block in our scheme. Through reflecting an input coherent
optical pulse from a cavity with a single trapped atom, we achieve a CPS
gate between the atom and the coherent optical pulse. Based on series of CPS
gates, an $n$-qubit HP gate is constructed in Sec. III. In Sec. IV, we show
how to realize the HP gate and use it for entanglement distribution in
detail. Finally, we end the paper with a discussion and conclusion in Sec. V.

\section{Basic Building Block}

The basic building block for our scheme is shown in Fig. {1(a)}, in which a
three-level atom is trapped in a single-side cavity. The qubit is
represented by different hyperfine levels $\left\vert 0\right\rangle $ and $%
\left\vert 1\right\rangle $ in the atomic ground-state manifold (see Fig. {%
1(b))}. The atomic transition $\left\vert 1\right\rangle \longleftrightarrow
\left\vert e\right\rangle $ is coupled to the cavity mode $a$ and driven by
the input field. While the state $\left\vert 0\right\rangle $ is decoupled
to the cavity mode due to the large hyperfine splitting.\ In the rotating
frame with respect to the frequency of input pulse, the interaction between
the atom and the cavity mode is described by the Hamiltonian

\begin{equation}
H=\delta _{1}\left\vert e\right\rangle \left\langle e\right\vert +\delta
_{2}a^{\dagger }a+g(a\sigma _{+}+a^{\dagger }\sigma _{-}),
\end{equation}%
where $\sigma _{+}=\left\vert e\right\rangle \left\langle 1\right\vert $, $%
\sigma _{-}=\left\vert 1\right\rangle \left\langle e\right\vert $, $\delta
_{1}=\omega _{0}-\omega _{p}$, $\delta _{2}=\omega _{c}-\omega _{p}$; $%
\omega _{0},$ $\omega _{c}$ and $\omega _{p}$ denote the resonant frequency
between the excited state $\left\vert e\right\rangle $ and the ground state $%
\left\vert 1\right\rangle $, the frequency of cavity mode and the input
pulse, respectively; $g$ is the atom-cavity coupling rate. The cavity mode $%
a $ is driven by the input field $a_{in}$. By omitting the Langevin noises
that have negligible contribution to the dynamics, we can easily obtain the
Heisenberg-Langevin equations for the internal cavity field and the atomic
operator

\begin{eqnarray}
&&\dot{a}=-\left( i\delta _{2}+\frac{\kappa }{2}\right) a-ig\sigma _{-}-%
\sqrt{\kappa }a_{in}  \notag \\
&&\dot{\sigma}_{-}=-\left( i\delta _{1}+\frac{\gamma }{2}\right) \sigma
_{-}+ig\sigma _{z}a
\end{eqnarray}%
where $\kappa $ and $\gamma $ are the cavity decay rate and the spontaneous
emission rate for the excited level, $\sigma _{z}=\left\vert e\right\rangle
\left\langle e\right\vert -\left\vert 1\right\rangle \left\langle
1\right\vert $, The output field $a_{out}$ is connected with the input by
the standard cavity input-output relation $a_{out}=a_{in}+\sqrt{\kappa }a.$
The input field has the standard commutation relations $[a_{in}(t),a_{in}^{%
\dagger }(t^{\prime })]=\delta (t-t^{\prime })$. Providing the cavity decay
rate $\kappa $ is sufficiently large, the atom will have a weak excitation.
So the component $\left\vert e\right\rangle \left\langle e\right\vert $ is
negligible. Under this approximation, $-\sigma _{z}$ can be replaced with
the population $P_{1}$ for the atom in the state $\left\vert 1\right\rangle $%
. If $\delta _{1}\gg \gamma /2$, from Eq. (2), we can derive the output field

\begin{equation}
a_{out}=ra_{in},
\end{equation}%
where the reflection coefficient is

\begin{equation}
r=\frac{i\delta _{1}(i\delta _{2}-\frac{\kappa }{2})+P_{1}g^{2}}{i\delta
_{1}(i\delta _{2}+\frac{\kappa }{2})+P_{1}g^{2}}.
\end{equation}

\begin{figure}[tbp]
\includegraphics[width=6cm]{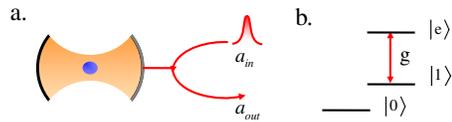}
\caption{(Color online) (a) Setup for the basic building block. An input
weak coherent optical pulse is reflected by a cavity with a single trapped
atom. (b) Level structure of the single-atom and the coupling induced by
cavity field.}
\end{figure}

Note that the population $P_{1}=0$ ($P_{1}=1$) for the atom in the state $%
\left\vert 0\right\rangle $ $(\left\vert 1\right\rangle ))$. We further
define $r_{0}$ and $r_{1}$ as the reflection coefficient corresponding to
the atomic states $\left\vert 0\right\rangle $ and $\left\vert
1\right\rangle $, and assume that the initial input optical pulse is
prepared in a coherent state $\left\vert \alpha \right\rangle $, and the
atom is initialized in a superposition state. The coherent state can be
written in the form of displacement operator, then the initial state of the
system is

\begin{eqnarray}
\left\vert \chi \right\rangle _{i} &=&\frac{\left\vert 0\right\rangle
+\left\vert 1\right\rangle }{\sqrt{2}}D(\alpha )\left\vert 0\right\rangle
_{in}  \notag \\
&=&\frac{\left\vert 0\right\rangle +\left\vert 1\right\rangle }{\sqrt{2}}%
e^{\alpha a_{in}^{\dagger }-\alpha ^{\ast }a_{in}}\left\vert 0\right\rangle
_{in}.
\end{eqnarray}%
After interacting with the cavity-atom system, the input optical pulse will
be transformed according to the reflection coefficient. That is, if the atom
is in the state $\left\vert 0\right\rangle $, the relationship between the
output field and the input field is $a_{out}=r_{0}a_{in}$ ($r_{0}$ is a
complex number), then we can get a cavity-assisted transformation for the
input coherent field with $a_{in}^{\dagger }\rightarrow
r_{0}a_{out}^{\dagger }$. While in the state $\left\vert 1\right\rangle $,
the cavity-assisted transformation for the input coherent field is $%
a_{in}^{\dagger }\rightarrow r_{1}a_{out}^{\dagger }$. By applying this
cavity-assisted transformations on the initial state $\left\vert \chi
\right\rangle _{i}$, one can get the final state

\begin{eqnarray}
\left\vert \chi \right\rangle _{f} &=&\frac{1}{\sqrt{2}}\left( \left\vert
0\right\rangle D(r_{0}\alpha )+\left\vert 1\right\rangle D(r_{1}\alpha
)\right) \left\vert 0\right\rangle _{out}  \notag \\
&=&\frac{1}{\sqrt{2}}\left( \left\vert 0\right\rangle \left\vert \alpha
e^{i\varphi _{0}}\right\rangle _{out}+\left\vert 1\right\rangle \left\vert
\alpha e^{i\varphi _{1}}\right\rangle _{out}\right) ,
\end{eqnarray}%
where $\varphi _{s}=\arg (r_{s})$, $s=0,1$. As we can see, with the
input-output process, a phase shift conditioned on the atomic state is
generated on the input coherent optical pulse, which yields a controlled
phase-shift (CPS) gate. This also extends the QC protocols {\color{blue}\cite%
{An,An2,Mei2} }with low-$Q$ cavity and single-photon pulse to continuous
variable regime. In the following section, based on series of CPS gates, we
will construct an $n$-qubit HP gate and use it for distributed quantum
computation.

\section{Hybrid Parity Gate}

In order to have a better understanding of the HP gate in our scheme, a
brief overview of the parity gate is given firstly. The parity gate has been
firstly proposed in linear-optic system to construct a controlled-NOT (CNOT)
gate probabilistically {\color{blue}\cite{Parity1}}. Soon afterwards
Beenakker and coauthors have shown that two charge parity gates with single
qubit gates can realize a CNOT gate deterministically {\color{blue}\cite%
{Parity2}}, which is a significant breakthrough for universal QC. These
achievements have stimulated great interest to implement a parity gate in
various systems for quantum information processing {\color{blue}\cite%
{Application}}. Now we give a brief introduction for the parity gate. It can
be viewed in the following implementation manner%
\begin{equation}
\left\vert xy\right\rangle \left\vert 0\right\rangle \rightarrow \left\vert
xy\right\rangle \left\vert x\oplus y\right\rangle ,
\end{equation}%
where $x$ and $y$ are the two input qubits, and an ancilla is introduced and
initialized firstly in the null state. The gate keeps the input state $%
\left\vert xy\right\rangle $ unchanged, but reveals the parity $\mathbf{p}%
=x\oplus y$ ($\oplus $ is the addition mod 2) by measuring the ancilla. If
the input state is in the product state $\left\vert +\right\rangle ^{\otimes
2}$, where $\left\vert +\right\rangle =$ $\left( \left\vert 0\right\rangle
+\left\vert 1\right\rangle \right) /\sqrt{2}$, the measurement result $%
\mathbf{p}=0$ projects the input state into an entangled state $\left(
\left\vert 00\right\rangle +\left\vert 11\right\rangle \right) /\sqrt{2}$,
and $\mathbf{p}=1$ into $\left( \left\vert 01\right\rangle +\left\vert
10\right\rangle \right) /\sqrt{2}$. A standard quantum network for the
parity gate has been studied in {\color{blue}\cite{Parity3}}, which uses a
qubit as an ancilla. Each input qubit interacts once with the ancillary
qubit via a CNOT gate. Finally the parity gate is achieved by measuring the
ancillary qubit state. This network could be carried out in a photonic
module {\color{blue}\cite{PMD}}.

Recently, Ionicioiu \textit{et al. }have generalized the above parity gate
to $n$-qubit case {\color{blue}\cite{Parity5}}. In their generalized parity
(GP) module, a qudit with $d$ dimensional Hilbert space serves as an
ancilla. Each qubit interacts once with the ancillary qudit by a controlled-$%
U$ gate, and $U$ $\left( \in U\left( d\right) \right) $ is a unitary
transformation. For different initial states of the qubits (i.e. with
different parities), after series of controlled-$U$ gates, the final qudit
states are orthonormal with each other. If the input state is $\otimes
_{i=1}^{n}\left\vert x_{i}\right\rangle $ ($x_{i}=0,1$), the generalized
parity could be defined as $\mathbf{p=}\sum\nolimits_{i=1}^{n}x_{i}$ mod $d$%
, which relies on the measurement outcome of the final qudit state. The
authors considered both the nondegenerate ($n=d$) and degenerate case ($n>d$%
) and showed how to generate entanglement in the two cases. The key demand
of this GP module is a controlled-$U$ gate between a qubit and a qudit. The
gate is also required to make the final qudit states orthonormal with each
other for different parities states of qubits. However, it is very difficult
to look for such qudit candidate in practice.

\begin{figure}[tbp]
\includegraphics[width=6cm]{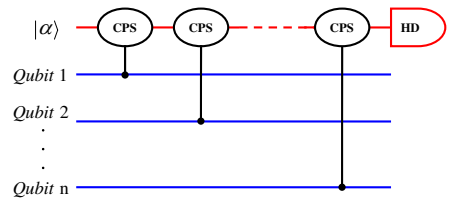}
\caption{(Color online) An $n$-qubit hybrid parity gate. HD is a homodye
detection on the final output phase-shift coherent state. }
\end{figure}

Here we further relax this constraint and employ the continuous quantum
variable instead of discrete variable as an ancilla. In our scheme, a
coherent optical pusle is used as an ancilla, and $n$ single atoms are the
input qubits. Each single atom is trapped in a cavity and interacts once
with the input coherent optical pulse by a CPS gate. As shown in Fig. 2, the
parity gate acts as%
\begin{equation}
\otimes _{i=1}^{n}\left\vert x_{i}\right\rangle \left\vert \alpha
\right\rangle \rightarrow \otimes _{i=1}^{n}\left\vert x_{i}\right\rangle
\left\vert \alpha \left( \mathbf{p}\right) \right\rangle ,
\end{equation}%
where $x_{i}$ is the $i$th input qubit, $\left\vert \alpha \right\rangle $
is the initial input ancillary coherent state, and $\left\vert \alpha \left(
\mathbf{p}\right) \right\rangle $ is the final output phase shift coherent
state. The parity here is defined as $\mathbf{p=}\sum%
\nolimits_{i=1}^{n}x_{i} $ mod $n$, which corresponds to the nondegenerate
case in the GP module {\color{blue}\cite{Parity5}}. As we will demonstrate
in detail in the next section, the phase shift of the final output coherent
state could reveal the parity of the initial input state. So through a
homodyne detection on the final output phase-shift coherent state, one can
get the parity and realize the HP gate.

\section{Quantum Entanglement Distribution}

In this section, we will show thoroughly how to realize an $n$-qubit HP gate
and use it for entanglement distribution among distant quantum nodes,
including $n$-qubit GHZ state $\left\vert GHZ_{n}\right\rangle $, W state $%
\left\vert W_{n}\right\rangle $, Dicke state $\left\vert
D_{n,k}\right\rangle $, and certain sums of Dicke states $\left\vert
G_{n,k}\right\rangle $. The entangled states $\left\vert
D_{n,k}\right\rangle $ and $\left\vert G_{n,k}\right\rangle $ have the
following forms%
\begin{eqnarray}
\left\vert D_{n,k}\right\rangle &=&\frac{1}{\sqrt{C_{n}^{k}}}\sum \tilde{P}%
_{k}\left( \left\vert 0\right\rangle ^{\otimes (n-k)}\left\vert
1\right\rangle ^{\otimes k}\right) ,  \notag \\
\left\vert G_{n,k}\right\rangle &=&\left\{
\begin{array}{cc}
\left( \left\vert D_{n,k}\right\rangle +\sigma _{x}^{\otimes n}\left\vert
D_{n,k}\right\rangle \right) /\sqrt{2}, & n\neq 2k; \\
\left\vert D_{2k,k}\right\rangle , & n=2k,%
\end{array}%
\right.
\end{eqnarray}%
where $C_{n}^{k}=n!/[k!(n-k)!]$, the sum in the Dicke state is over all $%
C_{n}^{k}$ permutations (produced by the permutation operator $\tilde{P}_{k}$%
) with $k$ qubits in the state $\left\vert 1\right\rangle $, $\ $and $\sigma
_{x}$ is the bit flip operation. Note that the W state $\left\vert
W_{n}\right\rangle $ is a special case of Dicke state $\left\vert
D_{n,k}\right\rangle $ with $k=1$. In the subsequent text, we will firstly
give some detailed examples for the distribution of all the above
entanglement with the HP gate, and then make a generalization to get the $n$%
-qubit HP gate.

\subsection{Two-qubit case}

\begin{figure}[tbp]
\includegraphics[width=7cm]{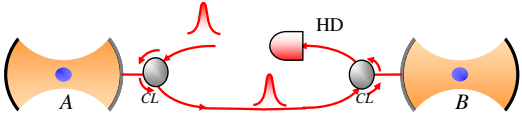}
\caption{(Color online) Setup for the two-qubit hybrid parity gate. $CL$ is
an optical circulator. }
\end{figure}

As shown in Fig. 3, two distant single atoms $A$ and $B$ are trapped in two
separated cavities. We assume the two single atoms are initially in the
product state $\left\vert \Psi _{2}\right\rangle _{i}=\left\vert
+\right\rangle ^{\otimes 2}$. If the input optical pulse is prepared in a
coherent state $\left\vert \alpha \right\rangle $, the initial state of the
whole system can be written as $\left\vert \Psi _{2}\right\rangle
_{i}\otimes \left\vert \alpha \right\rangle $. The input coherent optical
pulse interacts sequentially with the two distant atoms confined in the
respective cavities. It means that the coherent pulse undergoes two CPS
gates as shown in Fig. 2. Then the final state of the system is
\begin{eqnarray}
\left\vert \chi _{2}\right\rangle _{s} &=&\frac{1}{2}\left( \left\vert
01\right\rangle _{AB}+\left\vert 10\right\rangle _{AB}\right) \left\vert
\alpha e^{i\left( \varphi _{0}+\varphi _{1}\right) }\right\rangle +  \notag
\\
&&\frac{1}{2}\left( \left\vert 00\right\rangle _{AB}\left\vert \alpha
e^{2i\varphi _{0}}\right\rangle +\left\vert 11\right\rangle _{AB}\left\vert
\alpha e^{2i\varphi _{1}}\right\rangle \right) .
\end{eqnarray}%
In the case of low-$Q$ cavity ($\kappa \gg \gamma $), we set the phase shift
$\varphi _{0}=\pi /2$ and $\varphi _{1}=-\pi /2$ by adjusting $\delta
_{1}=\delta _{2}=\kappa /2$ and $g=\kappa /\sqrt{2}$. The state of the
system in this case becomes%
\begin{equation}
\left\vert \chi _{2}\right\rangle _{s}=\frac{1}{\sqrt{2}}\left( \left\vert
\phi ^{+}\right\rangle _{AB}\left\vert -\alpha \right\rangle +\left\vert
\psi ^{+}\right\rangle _{AB}\left\vert \alpha \right\rangle \right) ,
\end{equation}%
where $\left\vert \phi ^{+}\right\rangle _{AB}=\left( \left\vert
00\right\rangle _{AB}+\left\vert 11\right\rangle _{AB}\right) /\sqrt{2}$ and
$\left\vert \psi ^{+}\right\rangle _{AB}=\left( \left\vert 01\right\rangle
_{AB}+\left\vert 10\right\rangle _{AB}\right) /\sqrt{2}$ are the Bell
states. One can get the parity of the initial state of the atoms through
discriminating the two coherent states $\left\vert \alpha \right\rangle $
and $\left\vert -\alpha \right\rangle $, \ and project the two distant atoms
into a Bell state. Such discrimination can be achieved by a standard
homodyne detection {\color{blue}\cite{HD2,HD3,HD4}}. Here we choose a
homodyne measurement on the position quadrature of the coherent state $%
X=(a+a^{\dagger })/\sqrt{2}$. The wavefunction of coherent state in the
position space is

\begin{eqnarray}
\left\langle x|\alpha e^{i\theta }\right\rangle &=&\left( 1/\pi \right)
^{1/4}\exp [-(x-\sqrt{2}\cos \theta \alpha )^{2}/2+i\zeta \left( x,\theta
\right) ],  \notag \\
\zeta \left( x,\theta \right) &=&\left[ \alpha \sin \theta \left( x-2\alpha
\cos \theta \right) \right] \text{ mod }2\pi ,
\end{eqnarray}%
where $\alpha $ is real, $\left\vert x\right\rangle $ is an eigenstate of $X$
with eigenvalue $x$. Then the Eq. (11) can be rewritten as%
\begin{equation}
\left\vert \chi _{2}\right\rangle _{s}=\frac{1}{\sqrt{2}}\left( f(x,-\alpha
)\left\vert \phi ^{+}\right\rangle _{AB}+f(x,\alpha )\left\vert \psi
^{+}\right\rangle _{AB}\right) ,
\end{equation}%
where $f(x,\pm \alpha )=\left( 1/\pi \right) ^{1/4}\exp [-(x\mp \sqrt{2}%
\alpha )^{2}/2]$ are two Gaussian curves corresponding to the coherent state
$\left\vert \pm \alpha \right\rangle $. The midpoint of the two peaks of the
curves is located at $x_{m}=0$ and the distance between the two peaks is $%
x_{d}=2\sqrt{2}\alpha $. Upon the homodyne detection, we obtain the parity
and the final projected atomic entanglement as the following
\begin{eqnarray}
\text{ }\mathbf{p} &\mathbf{=}&1\text{ and }\left\vert \Psi
_{2}\right\rangle _{f}=\left\vert \psi ^{+}\right\rangle _{AB}\text{ for }%
x\geq x_{m}\text{;}  \notag \\
\mathbf{p} &\mathbf{=}&0\text{ and }\left\vert \Psi _{2}\right\rangle
_{f}=\left\vert \phi ^{+}\right\rangle _{AB}\text{ for }x<x_{m}\text{.}
\end{eqnarray}

\begin{figure}[tbp]
\includegraphics[width=8cm,height=6cm]{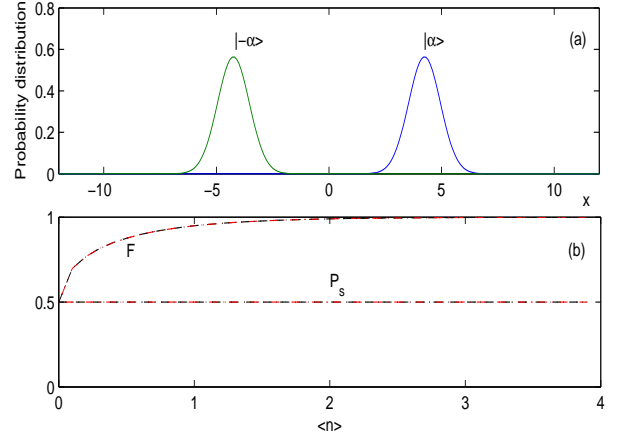}
\caption{(Color online) (a) The probability distribution for the final
output phase-shift coherent states $\left\vert \pm \protect\alpha %
\right\rangle $, with $\protect\alpha =3$. (b) shows the calculated success
probability and fidelity of the final projected atomic entanglement $%
\left\vert \protect\psi ^{+}\right\rangle _{AB}$ for the atomic spontaneous
rates $\protect\gamma =0$ (dashed curve), $\protect\gamma =0.2\protect\kappa
$ (dash-dot curve) and $\protect\gamma =0.5\protect\kappa $ (dotted curve).
Here we consider the channel transmission $\protect\eta ^{2}=2/3$.}
\end{figure}

As shown in Fig. 4(a), if the distance $x_{d}$ is not too small, one can
distinguish the two probability distributions with high fidelity by the
homodyne detection. However, there are some photon losses in practice, which
involve contributions from the channel attenuation, the nonideal cavity loss
and atomic spontaneous emission. For simplicity, the channel attenuation is
referred to as external loss, and the nonideal cavity loss and atomic
spontaneous emission are referred to as internal loss. These photon losses
will reduce the distinguishability in the homodyne detection. For
long-distance entanglement distribution in our scheme, the external loss is
the dominant photon loss and can be quantified with the parameter $1-\eta
^{2}$, where $\eta ^{2}$ is the efficient channel transmission. In the
following discussion, we neglect the internal loss first but come back to it
later using numerical simulation.

From the Eq. (14), the atoms will be projected into the Bell state $%
\left\vert \psi ^{+}\right\rangle _{AB}$ by the homodyne detection with the
measure result $x\geq x_{m}$. After considering the channel loss, the
success probability of such an event is
\begin{eqnarray}
P_{s} &=&\int_{x_{m}}^{+\propto }dx\text{ tr}_{atom}[\rho (x)]  \notag \\
&=&\frac{erfc(\sqrt{2}\eta \alpha )+erfc(-\sqrt{2}\eta \alpha )}{4}\text{,}
\end{eqnarray}%
where $\rho (x)$ is the density matrix of final state of the system, erfc is
the complementary error function. However, as shown in Fig. 4(a), the
probability distribution with respect to the coherent state $\left\vert
-\alpha \right\rangle $ also contributes to the measure result $x\geq x_{m}$%
, and it would have an influence on the fidelity of the projected Bell state
$\left\vert \psi ^{+}\right\rangle _{AB}$. Taking into account this
imperfection, the average fidelity is calculated as {\color{blue}\cite{CL}}

\begin{eqnarray}
F &=&\frac{1}{P_{s}}\int_{x_{m}}^{+\propto }dx_{AB}\left\langle \psi
^{+}\right\vert \rho (x)\left\vert \psi ^{+}\right\rangle _{AB}  \notag \\
&=&\frac{erfc(-\sqrt{2}\eta \alpha )}{erfc(\sqrt{2}\eta \alpha )+erfc(-\sqrt{%
2}\eta \alpha )}\text{.}
\end{eqnarray}%
With the channel transmission parameter $\eta ^{2}=2/3$, the success
probability and fidelity of the final projective state $\left\vert \psi
^{+}\right\rangle _{AB}$ are shown as a function of the mean photon number
of the input coherent optical pulse in Fig. 4(b). In the situation of
channel loss, the final phase-shift coherent state becomes $\left\vert \pm
\eta \alpha \right\rangle $ and their probabilities distribution are still
symmetrical around the midpoint $x_{m}=0$, hence the probabilities for the
above two measurement results are the same and equal to $1/2$, which
coincides with the numerical result in Fig. 4(b). The Fig. 4(b) also shows
that the fidelity increases with the mean photon number of the input
coherent state. Even in the presence of channel loss, one can get an ideal
fidelity of the projective entanglement by increasing $\alpha $. For the
initial input coherent optical pulse with $<n>=3$, a unit fidelity could be
achieved as shown in Fig. 4(b). Note that the two probability distribution
curves in Fig. 4(a) are highly symmetrical, so the success probability and
fidelity of the distributed entanglement $\left\vert \phi ^{+}\right\rangle
_{AB}$ is the same as $\left\vert \psi ^{+}\right\rangle _{AB}$.
Furthermore, we reconsider the atomic spontaneous $\gamma $ in Eq. (2) and
analyze numerically its influence on the fidelity and success probability of
the projected entanglement. The performance of the parity gate is calculated
with respect to $\gamma =0.2\kappa $ and $0.5\kappa$. The numerical result
shown in Figure. 4(b) demonstrates that the influence of the atomic
spontaneous emission on our scheme is very small, and the two-qubit parity
gate is still with high fidelity and high success probability. For the
nonideal cavity loss, it could be included through a modified atomic
spontaneous emission rate {\color{blue}\cite{DK}}.
\begin{figure}[t]
\includegraphics[width=8cm,height=8cm]{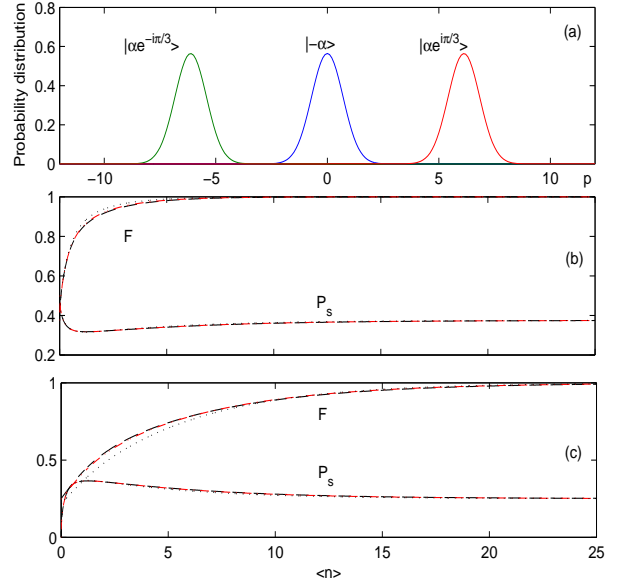}
\caption{(Color online) (a) The probability distribution for the final
output phase-shift coherent states $\left\vert -\protect\alpha \right\rangle
$ and $\left\vert \protect\alpha e\pm ^{i\protect\pi /3}\right\rangle $,
with $\protect\alpha =5$. (b) and (c) show the calculated success
probability and fidelity of the final projected atomic entanglement $%
\left\vert W_{3}\right\rangle $ $($or $D_{3,2})$ and $\left\vert
GHZ_{3}\right\rangle $ for the atomic spontaneous rates $\protect\gamma =0$
(dashed curve), $\protect\gamma =0.2\protect\kappa $ (dash-dot curve) and $%
\protect\gamma =0.5\protect\kappa $ (dotted curve), with the channel
transmission $\protect\eta ^{2}=2/3$. }
\end{figure}

\subsection{Three-qubit case}

For the three-qubit case, we will build a three-qubit HP gate and exploit it
for distributing various quantum entanglement, including the GHZ state, W
state and Dicke state. Like the configuration in Fig. 3, we assume three
distant atoms confined in three separated low-$Q$ cavities, and the atoms
are initially prepared in the product state $\left\vert \Psi
_{3}\right\rangle _{i}=$ $\left\vert +\right\rangle ^{\otimes 3}$. An input
coherent optical pulse experiences a CPS gate with each atom-cavity system
as illustrated in Fig. 2. We adjust $\delta _{1}=\delta _{2}=\sqrt{3}\kappa
/2$ and $g^{2}=3\kappa ^{2}/2$, and make the phase shift $\varphi _{0}=\pi
/3 $ and $\varphi _{1}=-\pi /3$ in the CPS gate, then the final state of the
system can be written as%
\begin{eqnarray}
\left\vert \chi _{3}\right\rangle _{s} &=&\frac{1}{2}\left\vert
GHZ_{3}\right\rangle \left\vert -\alpha \right\rangle +\frac{\sqrt{6}}{4}%
\left\vert W_{3}\right\rangle \left\vert \alpha e^{i\pi /3}\right\rangle
\notag \\
&&+\frac{\sqrt{6}}{4}\left\vert D_{3,2}\right\rangle \left\vert \alpha
e^{-i\pi /3}\right\rangle .
\end{eqnarray}
In order to completely distinguish the above output phase-shift coherent
states, we adopt the homodyne detection on the momentum quadrature of the
coherent state $P=(a-a^{\dagger })/\sqrt{2}i$. Using the wavefunction of the
coherent state in the momentum space
\begin{eqnarray}
\left\langle p|\alpha e^{i\theta }\right\rangle &=&\left( 1/\pi \right)
^{1/4}\exp [-(p-\sqrt{2}\sin \theta \alpha )^{2}/2+i\zeta \left( p,\theta
\right) ],  \notag \\
\zeta \left( p,\theta \right) &=&\left[ -2\alpha \cos \theta (\sqrt{2}%
p-\alpha \sin \theta )\right] \text{ mod }2\pi ,
\end{eqnarray}
where $\alpha $ is real, $\left\vert p\right\rangle $ is an eigenstate of $P$
with eigenvalue $p$, the final state of the system can be rewritten as

\begin{eqnarray}
\left\vert \chi _{3}\right\rangle _{s} &=&\frac{1}{2}f\left( p,0\right)
e^{i\zeta \left( p,\pi \right) }\left\vert GHZ_{3}\right\rangle  \notag \\
&&+\frac{\sqrt{6}}{4}f\left( p,\sqrt{3}\alpha /2\right) e^{i\zeta \left(
p,\pi /3\right) }\left\vert W_{3}\right\rangle  \notag \\
&&+\frac{\sqrt{6}}{4}f\left( p,-\sqrt{3}\alpha /2\right) e^{i\zeta \left(
p,-\pi /3\right) }\left\vert D_{3,2}\right\rangle ,
\end{eqnarray}
where $f\left( p,\beta \right) =$ $\left( 1/\pi \right) ^{1/4}\exp [-(p-%
\sqrt{2}\beta )^{2}/2]$, $\beta =0$, $\pm \sqrt{3}\alpha /2$. In Fig. 5(a),
we have plotted the probability distribution for the output phase-shift
coherent state $\left\vert \alpha e^{i\theta }\right\rangle $, $\theta =\pi
, $ $\pm \pi /3$. The midpoints between the peaks of the probability
distribution for $\left\vert \alpha e^{i\pm \pi /3}\right\rangle $ and for $%
\left\vert \alpha e^{i\pi }\right\rangle $ are located at $p_{m\pm }=\pm
\sqrt{6}\alpha /4$. After the homodyne detection, the parity and final
projected atomic entanglement are

\begin{eqnarray}
\text{\ }\mathbf{p} &\mathbf{=}&1\text{ and }\left\vert \Psi
_{3}\right\rangle _{f}=\left\vert W_{3}\right\rangle \text{ for }p>p_{m+};
\notag \\
\mathbf{p} &\mathbf{=}&0\text{ and }\left\vert \Psi _{3}\right\rangle
_{f}=\left\vert GHZ_{3}\right\rangle \text{ for }p_{m-}\leq p\leq p_{m+};
\notag \\
\mathbf{p} &\mathbf{=}&2\text{ and }\left\vert \Psi _{3}\right\rangle
_{f}=\left\vert D_{3,2}\right\rangle \text{ for }p<p_{m-}\text{.}
\end{eqnarray}%
As the two-qubit case, we also consider the influence of the channel loss
and calculate the success probability and fidelity. The result is shown in
Fig. 5(b) and (c). When the amplitude of the input coherent state $\alpha =5$%
, as one would expect from the Eq. (19), the success probability for the
final projected entanglement $\left\vert W_{3}\right\rangle $ or $\left\vert
D_{3,2}\right\rangle $ is $P_{s}=3/8$, for $\left\vert GHZ_{3}\right\rangle $
is $P_{s}=1/4$, and all the fidelities can approach unit. The figure 5(b)
and (c) also show that the entanglement distribution can still be achieved
with high fidelity and high success probability even in the presence of the
atomic spontaneous emission.

\subsection{Distribution of sums of Dicke state}

The sum of Dicke states $\left\vert G_{n,k}\right\rangle $ has been recently
introduced in {\color{blue}\cite{QSS2} }for quantum secret sharing. The most
important step for this purpose is the distribution of quantum entanglement
among distant parties. For a particular case, the distribution of $%
\left\vert G_{3,1}\right\rangle $ is shown in the next. Instead of $P$, the
homodyne measurement on the position quadrature $X$ of the final output
coherent state in Eq. (17) is applied. After such measurement, the state of
the system becomes

\begin{equation}
\left\vert \chi _{3}\right\rangle _{s}=\frac{1}{2}\left( f\left( x,-\alpha
\right) \left\vert GHZ_{3}\right\rangle +\sqrt{3}f\left( x,\alpha /2\right)
\left\vert G_{3,1}^{\prime }\right\rangle \right) ,
\end{equation}%
where $f(x,\beta )=\left( 1/\pi \right) ^{1/4}\exp [-(x-\sqrt{2}\beta
)^{2}/2]$, $\beta =-\alpha ,\alpha /2$, $\left\vert G_{3,1}^{\prime
}\right\rangle =\left[ e^{i\zeta \left( x,\pi /3\right) }\left\vert
D_{3,1}\right\rangle +e^{-i\zeta \left( x,\pi /3\right) }\left\vert
D_{3,2}\right\rangle \right] /\sqrt{2}$. We have plotted the probability
distribution for the output phase-shift coherent states in Fig. 6(a). The
probability distribution for the output coherent states $\left\vert \alpha
e^{\pm i\pi /3}\right\rangle $ in the position space are the same, so one
cannot determine which Dicke entanglement is projected. This yields the sum
of Dicke states $\left\vert G_{3,1}^{\prime }\right\rangle $, which can be
transformed into the standard form $\left\vert G_{3,1}\right\rangle $ as
shown in Eq. (9) through local unitary operations. The midpoint between the
peaks of the probability distribution for $\left\vert \alpha e^{\pm i\pi
/3}\right\rangle $ and $\left\vert \alpha e^{i\pi }\right\rangle $ is
located at $x_{m}=-\sqrt{2}\alpha /4$.\ Thus the projected atomic entangled
states are

\begin{eqnarray}
\left\vert \Psi _{3}\right\rangle _{f} &=&\left\vert G_{3,1}^{\prime
}\right\rangle \text{ for }x\geq x_{m};  \notag \\
\left\vert \Psi _{3}\right\rangle _{f} &=&\left\vert GHZ_{3}\right\rangle
\text{ for }x<x_{m}\text{.}
\end{eqnarray}%
The success probability and fidelity for the two projected entanglement are
plotted in Fig. 6(b) and (c). It is found that, only using an initial input
coherent state with $<n>=5$, the success probability and fidelity for the
final projected sum of Dicke state or GHZ state could be the ideal case as
predicted in Eq. (21), i.e. $P_{s}\left( \left\vert G_{3,1}^{\prime
}\right\rangle \right) =3/4$, $P_{s}\left( GHZ_{3}\right) =1/4$, and $%
F\left( GHZ_{3}\right) =F\left( \left\vert G_{3,1}^{\prime }\right\rangle
\right) =1$. The numerical results in Fig. 6(b) and (c) also indicate that
the the atomic spontaneous emission has no significant effect on the
performance of the parity gate and one can complete entanglement
distribution with high fidelity and high success probability.

\begin{figure}[h]
\includegraphics[width=8cm,height=8cm]{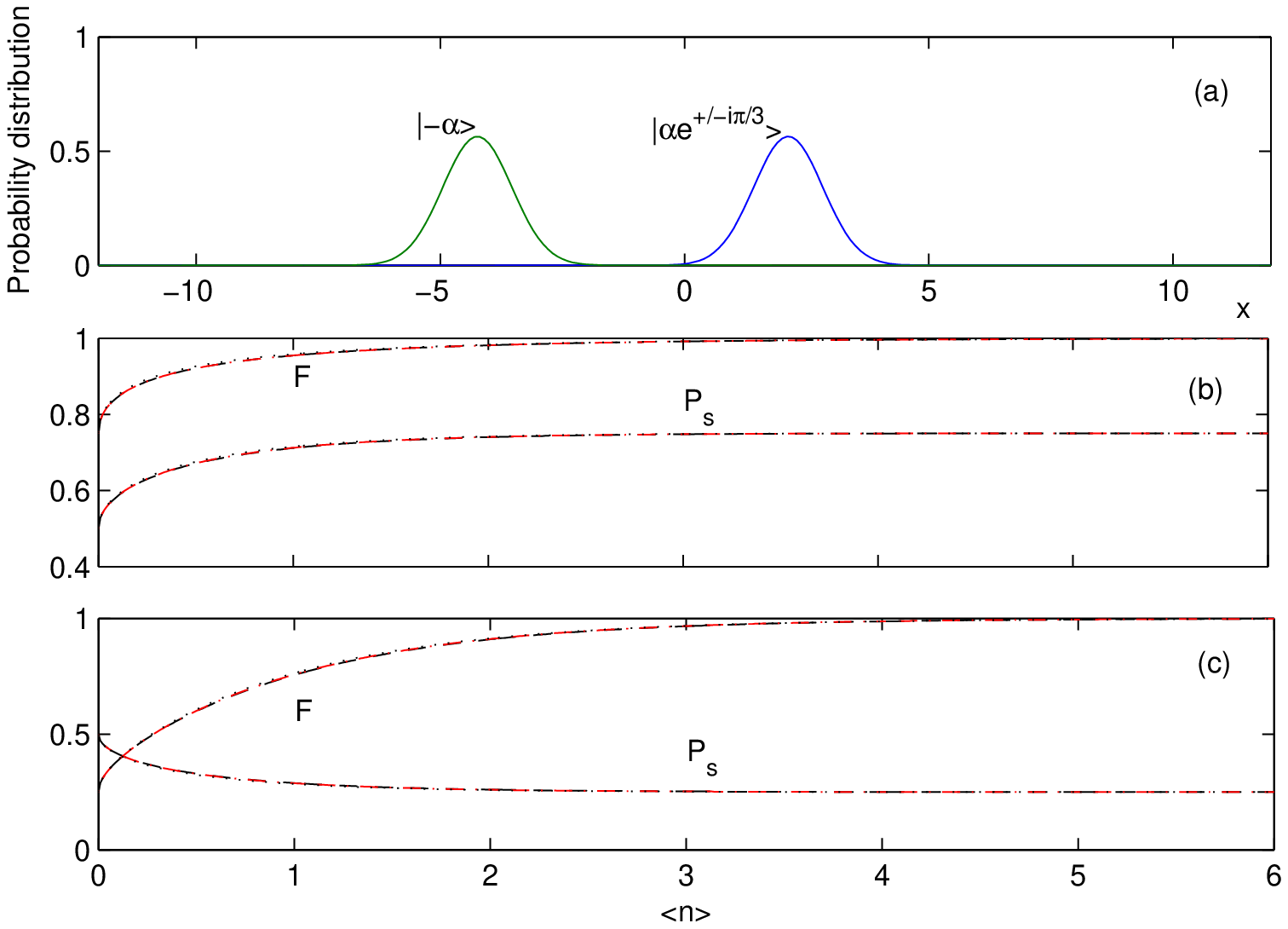}
\caption{(Color online) (a) The probability distribution for the final
output phase-shift coherent states $\left\vert -\protect\alpha \right\rangle
$ and $\left\vert \protect\alpha e\pm ^{i\protect\pi /3}\right\rangle $,
with $\protect\alpha =3$. (b) and \ (c) show the calculated success
probability and fidelity of the final projected atomic entanglement $%
\left\vert G_{3,1}^{\prime }\right\rangle $ and $\left\vert
GHZ_{3}\right\rangle $ for the atomic spontaneous rates $\protect\gamma =0$
(dashed curve), $\protect\gamma =0.2\protect\kappa $ (dash-dot curve) and $%
\protect\gamma =0.5\protect\kappa $ (dotted curve), with the channel
transmission $\protect\eta ^{2}=2/3$.}
\end{figure}

\subsection{$n$-qubit generalization}

According to the above detailed examples, we make a brief generalization to
construct the $n$-qubit HP gate and show its use for entanglement
distribution. Firstly, we assume $n$ single atoms trapped in separated
cavities are initially prepared in the state $\left\vert +\right\rangle
^{\otimes n}$. An input coherent optical pulse, which is initiated in the
coherent state $\left\vert \alpha \right\rangle $, interacts with each atom
by a CPS gate. In the CPS gate, we set the phase shift $\varphi _{0}=\pi /n$
and $\varphi _{1}=-\pi /n$ by adjusting the parameters of the cavity-atom
system. The final state of the system can be written as

\begin{eqnarray}
\left\vert \chi _{n}\right\rangle _{s} &=&\frac{1}{\sqrt{2^{n}}}\left( \sqrt{%
2}\left\vert GHZ_{n}\right\rangle \left\vert -\alpha \right\rangle +\right.
\notag \\
&&\left. \sum\nolimits_{k=1}^{n-1}\sqrt{C_{n}^{k}}\left\vert
D_{n,k}\right\rangle \left\vert \alpha e^{i(1-2k/n)\pi }\right\rangle
\right) .
\end{eqnarray}%
Now if we perform the homodyne detection on the momentum quadrature $P$ of
the final output phase-shift coherent state, the $n$ atoms would be
projected into a GHZ state $\left\vert GHZ_{n}\right\rangle $ $\left(
\mathbf{p=}0\right) $ or Dicke state $\left\vert D_{n,k}\right\rangle $ $%
\left( \mathbf{p=}k\right) $. Note that for $n=2k$, one needs an additional
HP gate with homodyne detection on the position quadrature $X$ to
distinguish the symmetric Dicke state $\left\vert D_{2k,k}\right\rangle $
and the GHZ state $\left\vert GHZ_{2k}\right\rangle $. Besides, if we
initially perform a homodyne detection on the position quadrature $X$ of the
output phase-shift coherent state in Eq. (23), the $n$ atoms would be
projected into a GHZ state $\left\vert GHZ_{n}\right\rangle $ or certain
sums of Dicke states $\left\vert G_{n,k}\right\rangle $ which could be used
for quantum secret sharing {\color{blue}\cite{QSS2}}. Because $\left\vert
W_{n}\right\rangle =\left\vert D_{n,1}\right\rangle =\sigma _{x}^{\otimes
n}\left\vert D_{n,n-1}\right\rangle $, the success probability for preparing
W state in our scheme is $P_{s}=n/2^{n-1}$, which keeps an exponential gain
over the scheme in {\color{blue}\cite{Wstate}}.

As we know, the GHZ state and Dicke state (including the W state) are not
equivalent under stochastic local operations and classical communication
(SLOCC) {\color{blue}\cite{SLOCC}}. This classification is particularly
relevant for evaluating the use of states for multiparty quantum
communication since states of the same SLOCC class can be used for the same
applications. Several of two inequivalent entangled states above have been
experimentally observed, like the GHZ state $\left\vert GHZ_{6}\right\rangle
$ {\color{blue}\cite{GHZ6} }and W state $\left\vert W_{8}\right\rangle $ {%
\color{blue}\cite{W8} }in ion trap system, and the Dicke state $\left\vert
D_{4,2}\right\rangle $ {\color{blue}\cite{D1}}, $\left\vert
D_{6,3}\right\rangle $ {\color{blue}\cite{D2}} in linear-optical system.
However, all these experiments are tailored to a particular entanglement,
and can not provide a choice between different SLOCC inequivalent
entanglement. In contrast, our scheme here gives a flexible physical way for
preparing SLOCC inequivalent entanglement, which has also been presented
recently in linear-optical system {\color{blue}\cite{D3}}.

\section{Discussion and Conclusion}

Before ending the paper, we give a brief discussion on the requirement of
atom-cavity system in the two- and three-qubit parity gate. First, in either
case, the detuning of the input coherent optical pulse with respect to the
atomic resonance and the cavity mode $\delta _{1}$ and $\delta _{2}$ are
required to be the same. This could be satisfied by employing a resonant
interaction between the atom and the cavity mode with $\omega _{c}=\omega
_{0}$. Then one can change the frequency of the input coherent light $\omega
_{p}$ to achieve the desired detuning. Meanwhile, the atom-cavity coupling
strength should be a required rate. For a Fabry-Perot cavity, the actual
atom-cavity coupling strength depends on the atomic position through $%
g\left( \mathbf{r}\right) =g_{0}\cos \left( k_{c}z\right) \exp \left[
-r_{\perp }^{2}/w_{c}^{2}\right] $, where $g_{0}$ is the peak coupling rate,
$r_{\perp }$ is the radial distance of the atoms with respect to the cavity
axis, $w_{c}$ and $k_{c}$ are the width and the wave vector of the Gaussian
cavity mode. Recent experiments have demonstrated the ability to manipulate
the position of a single atom {\color{blue}\cite{SA1,SA2} }or a BEC {%
\color{blue}\cite{BEC} }relative to the cavity mode using an atomic
conveyor. With this experimental technology, the atom-cavity coupling
strength can be tuned to controll the reflectivity of the input coherent
optical pulse. Note that such experiment has been performed recently using a
quantum dot confined in a photonic crystal cavity {\color{blue}\cite{CR}}.
Moreover, the similar controlled phase shifts up to $\pi /4$ have also been
observed in the same system {\color{blue}\cite{Shift}}. So our scheme fits
the current experimental technologies very well and could be feasible for
large-scale quantum computation and quantum communication in the future.

In conclusion, with single atoms trapped in separated low-$Q$ cavities
working in the intermediate coupling region as quantum nodes and a coherent
optical pulse as quantum channel, we have proposed a scheme for atomic
entanglement distribution over different nodes using a HP gate. In our
scheme, through an input coherent optical pulse reflecting from a quantum
node, a CPS gate between the atom and the coherent pulse is generated. With
the CPS gate and homodyne detection, we have shown how to construct an $n$%
-qubit HP gate, and found that the HP gate is very flexible for preparing
and distributing various interesting entanglement among different nodes,
like GHZ state $\left\vert GHZ_{n}\right\rangle $, W state $\left\vert
W_{n}\right\rangle $, Dicke state $\left\vert D_{n,k}\right\rangle $ and
certain sums of Dicke states $\left\vert G_{n,k}\right\rangle $. Our scheme
is within the present-day experiment technology as it does not need strong
coupling cavity and single-photon source. Even in the presence of channel
loss and atomic spontaneous emission, the scheme can work with high success
probability and high fidelity. In addition, the HP gate here has also an
inherent distribution feature, it may function as a basic building block for
a distributed QIP network.

\section{Acknowledgment}

This work is supported by the National Natural Science Foundation of China
under Grant No. 60978009, the National Basic Research Program of China
("973" Program) (2007CB925204 and 2009CB929604), and NUS Academic Research
(Grant No. WBS: R-144-000-189-305).

\end{document}